\documentstyle[11pt]{article}
\newcommand{\beq}{\begin{equation}}
\newcommand{\eeq}{\end{equation}}
\newcommand{\bea}{\begin{eqnarray}}
\newcommand{\eea}{\end{eqnarray}}
\begin{document}
\thispagestyle{empty}
\vspace{3cm}
%\begin{center}
\title{\Large{\bf Gauge Independence of the Lagrangian Path Integral \\in a
Higher-Order Formalism} }
%\end{center}
\vspace{2cm}
%\begin{center}
\author{{\sc I.A. Batalin}\\I.E. Tamm Theory Division\\
P.N. Lebedev Physics Institute\\Russian Academy of Sciences\\
53 Leniniski Prospect\\Moscow 117924\\Russia\\
~\\~{\sc K. Bering}\\Institute of Theoretical Physics\\
Uppsala University\\P.O. Box 803\\S-751 08 Uppsala\\Sweden\\~\\and\\~\\
{\sc P.H. Damgaard}\\The Niels Bohr Institute\\Blegdamsvej 17\\
Dk-2100 Copenhagen\\Denmark}
%\end{center}
\maketitle
\begin{abstract}
We propose a Lagrangian path integral based on gauge symmetries
generated by a symmetric higher-order $\Delta$-operator, and 
demonstrate that
this path integral is independent of the chosen gauge-fixing function.
No explicit change of variables in the functional integral is required 
to show this.
\end{abstract}
\vspace{4cm}
\begin{flushleft}
NBI-HE-96-47\\UUITP-20/96\\hep-th/9609037
\end{flushleft}

\vfill
\newpage

\noindent
{\em 1.\ Introduction.}~
The usual field-antifield formalism for gauge theories \cite{BV} is based
on a Grassmann-odd nilpotent operator $\Delta$, which
is assumed to be of 2nd order. In covariant form \cite{cov},
\beq
\Delta = \frac{1}{2} (-1)^{\epsilon_A}\frac{1}{\rho(\Gamma)}
\partial_A \rho(\Gamma) E^{AB}(\Gamma)\partial_B ~.
\label{Delclas}
\eeq
Here $\rho(\Gamma)$ coincides with the measure density in the functional
integral, and the non-degenerate $E^{AB}$ satisfies the following 
symmetry condition:
\beq
E^{AB} = -(-1)^{(\epsilon_A+1)(\epsilon_B+1)}E^{BA} ~.
\label{Esym}
\eeq

The $\Delta$-operator is the basic building block of the whole
Lagrangian quantization program in the field-antifield formalism. 
{}From it derives the Master Equations as well as the antibracket,
which in turn provides an anticanonical framework on the space
of fields and antifields. The $\Delta$-operator also
generates a global symmetry which is useful for
proving independence of the chosen gauge-fixing surface on the 
field-antifield supermanifold. 

In view of this, it is natural to ask if an analogous 
Lagrangian quantization 
prescription can be established with the help of a $\Delta$-operator
of arbitrary (perhaps even infinite) order. Investigations along
these lines were initiated by a study of possible quantum
deformations of the field-antifield formalism \cite{BT}. More recently,
such considerations have been prompted by the observed analogies 
between the $\Delta$-operator and a certain quantized 
Hamiltonian BRST operator $\Omega$ \cite{BF}
in the ghost momentum representation \cite{AD}, as well as
by a study of the associated algebraic structure \cite{BDA}. 
The relevant mathematical foundations date back to work of Koszul \cite{KA}.

\vspace{0.7cm}

\noindent
{\em 2.\ Gauge Independence.}~
Let there be given a functional measure $d\mu \equiv d\Gamma d\lambda
\rho(\Gamma)$. While $\Gamma$ represents the usual set of fields and
antifields, the $\lambda$'s can be seen either as implementing
the gauge (or hypergauge) fixing condition \cite{BT0}, or as the ghost 
fields for which the antifields are usual antighosts \cite{AD1}.
We assume that the general odd $\Delta$-operator satisfies two properties: 
It is nilpotent, 
\beq
\Delta^2 = 0 ~,
\label{Delnil}
\eeq
and it is {\em symmetric},
\beq
\Delta^T = \Delta~.
\label{Delsym}
\eeq
Here the transposed operator $\Delta^T$ is defined by
\beq
\int d\mu ~F \Delta G ~=~ (-1)^{\epsilon_{F}}\int d\mu \left(\Delta^T F
\right) G ~.
\eeq
It is conventional to assume in addition that
$\Delta(1) = 0$, but this assumption can easily be discarded \cite{KA,BDA},
and in fact we do not need it here.

Consider the two Master Equations
\beq
\Delta e^{\frac{i}{\hbar}W} = 0 ~~,~~~~~~~\Delta e^{\frac{i}{\hbar}X} = 0 ~,
\eeq
and the path integral \cite{BT0,BMS}
\beq
Z_X ~\equiv~ \int d\mu ~e^{\frac{i}{\hbar}[W+X]} ~.
\eeq

We wish to establish that $Z_X = Z_{X'}$ for an {\em arbitrary} deformation
$X \to X'$ that preserves the master equation for $X$. We consider here
$\Delta$-operators and solutions $X$ belonging to the class for which the following represents a maximal deformation \cite{BT0,BMS}:
\beq
e^{\frac{i}{\hbar}X'} = e^{[\Delta,\Psi]}e^{\frac{i}{\hbar}X} ~,
\label{maxdeform}
\eeq
i.e.\ for an infinitesimal transformation,
\beq
\delta e^{\frac{i}{\hbar}X} = [\Delta,\Psi]e^{\frac{i}{\hbar}X} ~.
\label{infXdef}
\eeq

Proof: ~We will use 3 ingredients: 1) The Master Equation for $W$, 
~2) The Master Equation for $X$, and ~3) The symmetry of $\Delta$.
Then,
\bea
Z_{X'} - Z_{X} &=& \int d\mu ~e^{\frac{i}{\hbar}W} [\Delta,\Psi]
e^{\frac{i}{\hbar}X} \cr
&=& \int d\mu \left[\left(\Delta e^{\frac{i}{\hbar}W}\right)\Psi 
e^{\frac{i}{\hbar}X} +
e^{\frac{i}{\hbar}W}\Psi\Delta e^{\frac{i}{\hbar}X}\right] \cr
&=& 0 ~.
\label{Zindep}
\eea

\vspace{0.7cm}
\noindent
{\em 3.\ BRST Symmetry.}~
We can also understand gauge independence of the path integral
from the existence of a nilpotent BRST symmetry. We can define two 
BRST transformations,
one associated with $W$, and one associated with $X$ \cite{AD,BDA}. 
They are:\footnote{We can also consider
their difference \cite{BMS}, or in fact any linear combination. 
When both Master Equations are satisfied, we can
remove the commutator, but the present form is convenient.}
\bea
\sigma_W F &~=~& \left({\scriptstyle \frac{\hbar}{i}}\right) e^{-\frac{i}{\hbar}W}[\Delta,F]
e^{\frac{i}{\hbar}W} \cr
\sigma_X F &~=~&   \left({\scriptstyle \frac{\hbar}{i}}\right) e^{-\frac{i}{\hbar}X}[\Delta,F]
e^{\frac{i}{\hbar}X} ~.
\label{sigmaop}
\eea
The Master Equations for $W$ and $X$ are preserved under transformations
$W \to W +\sigma_W\Psi$ and $X \to X + \sigma_X\Psi$. Thus another
way of phrasing the gauge independence of the path integral is
$Z_X = Z_{X+\delta X}$ with $\delta X \equiv \sigma_X\Psi$. This is
precisely the content of eqs. (\ref{infXdef}) and (\ref{Zindep}).
The operators (\ref{sigmaop}) are the natural generalizations of the
so-called quantum BRST operator for the case of the conventional
2nd order $\Delta$-operator \cite{H}.

If we define BRST invariant operators $G$ by
\beq
\sigma_W G ~=~ 0 ~,
\eeq
we observe that expectation values of such operators do not depend on $X$:
\bea
\langle G \rangle_{X+\delta X} - \langle G \rangle_X &=&
Z^{-1}\int d\mu ~G e^{\frac{i}{\hbar}[W+X]} e^{-\frac{i}{\hbar}X}[\Delta,\Psi]e^{\frac{i}{\hbar}X} \cr
&=& (-1)^{\epsilon_{G}}Z^{-1}\int d\mu ~\Delta\left(G e^{\frac{i}{\hbar}W}
\right) \Psi e^{\frac{i}{\hbar}X} \cr
&=& (-1)^{\epsilon_{G}}Z^{-1}\int d\mu \left([\Delta,G]e^{\frac{i}
{\hbar}W}\right) \Psi e^{\frac{i}{\hbar}X} \cr
&=& (-1)^{\epsilon_{G}}Z^{-1}\int d\mu ~e^{\frac{i}{\hbar}W} 
\left(  {\scriptstyle \frac{i}{\hbar}} \sigma_W G\right) \Psi e^{\frac{i}{\hbar}X} \cr
&=& 0 ~.
\eea

Similarly we can show that $\langle \sigma_W F \rangle = 0$ for any $F$:
\bea
\langle \sigma_W F \rangle &=& Z^{-1}\int d\mu ~e^{\frac{i}{\hbar}[W+X]}
\sigma_W F \cr
&=& Z^{-1}\int d\mu ~e^{\frac{i}{\hbar}[W+X]}  \left({\scriptstyle
\frac{\hbar}{i}}\right) 
e^{-\frac{i}{\hbar}W} [\Delta,F] e^{\frac{i}{\hbar}W} \cr
&=& Z^{-1}  \left({\scriptstyle \frac{\hbar}{i}}\right)  
\int d\mu \left[\left(\Delta e^{\frac{i}{\hbar}X}
\right)Fe^{\frac{i}{\hbar}W} - (-1)^{\epsilon_F}
e^{\frac{i}{\hbar}X} F \Delta e^{\frac{i}{\hbar}W}\right] \cr
&=& 0 ~.
\eea
The formalism is symmetric under exchanges of $W$ and $X$. The
choice of boundary conditions stipulates which part will play the
r\^{o}le of action (here taken to be $W$), and which part will play
the r\^{o}le of gauge fixing (here taken to be $X$).

\vspace{0.7cm}
\noindent
{\em 4.\ Generalizations.}~
We note that the requirement of symmetry of $\Delta$, eq. (\ref{Delsym}), can 
be relaxed if we instead impose two conjugate Master Equations on $W$ and
$X$:
\beq
\Delta e^{\frac{i}{\hbar}W} = 0 ~~,~~~~~~~
\Delta^T e^{\frac{i}{\hbar}X} = 0 ~.
\eeq
Under arbitrary infinitesimal deformations
\beq 
\delta e^{\frac{i}{\hbar}X} = [\Delta^T,\Psi]e^{\frac{i}{\hbar}X} ~,
\eeq
one finds again $Z_{X} = Z_{X'}$. Similarly the BRST operator $\sigma_X$
of eq. (\ref{sigmaop}) will have $\Delta$ replaced by $\Delta^T$. The rest
of the conclusions then remain unaltered.  

\vspace{0.7cm}
\noindent
{\em 5.\ Quantum Deformations of  $\Delta$.}~
Up to this point we have made no further assumptions about $\Delta$
beyond those stated in eq. (\ref{Delnil}-\ref{Delsym}), and indirectly
in eq. (\ref{maxdeform}). 
Gauge independence of the proposed path
integral (\ref{Zindep}) holds in all generality. 
We shall end by some comments specific
to $\Delta$-operators obtained by quantum deformations of the classical
2nd order $\Delta$-operator (\ref{Delclas}) \cite{BT}. 
Quantum corrections are expected to arise when operator-ordering
in the Hamiltonian formalism is properly taken into account. 

The most  general quantum deformation of the 2nd order $\Delta$ 
can be written in terms of its homogeneous components as 
\bea
\Delta &=& \sum_{n=0}^{\infty} \left( {\scriptstyle \frac{\hbar}{i}}  
\right)^n \Delta_n \cr
\Delta_n = \sum_{m=0}^{n+2} \Delta_{n,m}~~, ~&&~~~
\Delta_{n,m} = \Delta^{A_{m}\ldots A_{1}}_{n,m}(\Gamma) \
\partial_{A_{1}}\ldots \partial_{A_{m}} ~.
\label{deltaexpand}
\eea
in an ordering with all (left) derivatives standing to the right.     
$\left( {\scriptstyle \frac{\hbar}{i}} \right)^n\Delta_{n,m}$ is the 
contribution to $\Delta$ of order $n$ in  $\hbar$ 
and differential order $m$. So the classical part is\footnote{The original
$\Delta$-operator (\ref{Delclas}) is of this form, with $\Delta_{0,0} 
= 0$.}
\beq
 \Delta_0 = \Delta_{0,2} +  \Delta_{0,1} + \Delta_{0,0} ~.
\eeq
Each new order of $\hbar$ gives rise to one extra order 
of differentiation.

The Master Equation has an expansion
in terms of higher antibrackets $\Phi^k$ \cite{AD,BDA}:
\bea
 0 &=&  \left( {\scriptstyle \frac{\hbar}{i}}  \right)^2 
e^{-\frac{i}{\hbar}W}\Delta e^{\frac{i}{\hbar}W} \cr
 &=&  \left( {\scriptstyle \frac{\hbar}{i}}  \right)^2  \sum_{k=0}^{\infty} \frac{1}{k!}  
\Phi^k_{\Delta} \left( {\scriptstyle \frac{i}{\hbar}} W, \ldots, 
{\scriptstyle \frac{i}{\hbar}} W \right)  \cr
&=&  \sum_{n=0}^{\infty} \sum_{k=0}^{n+2} 
\left( {\scriptstyle \frac{\hbar}{i}}  \right)^{n+2-k} \frac{1}{k!}
\Phi^k_{\Delta_n} \left( W, \ldots, W \right)  \cr
&=&  \sum_{k=2}^{\infty}\frac{1}{k!} \Phi^k_{\Delta_{k-2}} \left( W, \ldots, 
W \right) 
 + {\scriptstyle \frac{\hbar}{i}} \sum_{k=1}^{\infty} \frac{1}{k!}
\Phi^k_{\Delta_{k-1}} \left( W, \ldots, W \right)  \cr
&&+ \sum_{\ell=2}^{\infty} 
 \left( {\scriptstyle \frac{\hbar}{i}}  \right)^{\ell} \sum_{k=0}^{\infty}\frac{1}{k!}
\Phi^k_{\Delta_{k+\ell-2}} \left( W, \ldots, W \right) ~.
\eea
Here we have introduced the generalized antibracket \cite{KA}
\beq
\Phi^k_{\Delta} \left( A_1, \ldots, A_k \right) 
= \left[ \cdots  \left[ \Delta ,  A_1 \right], \cdots , A_k \right]1 ~,
\eeq
in terms of $k$ nested commutators.

Assuming a semiclassical expansion of $W$, ~$W = \sum_{n=0}^{\infty}
\hbar^n W_n$, one sees that the classical master equation 
\beq
\Phi_{\Delta_0}^2(W_0,W_0) = 0
\label{oriclasmast}
\eeq 
will be replaced, in general, by a (possibly infinite\footnote{All sums over
the order of the bracket $k$ truncate at order $N$ if the $\Delta$-operator
is of order $N$.}) sum,
\beq
  \sum_{k=2}^{\infty} \frac{1}{k!} \Phi^k_{\Delta_{k-2}} 
\left(W_0, \ldots, W_0 \right) = 0 ~.
\label{defclasmast}
\eeq
A similar analysis for the operator $\sigma_W$ yields
\beq
\sigma_W F = \sum_{k=1}^{\infty}\frac{1}{k!}
\Phi^{k+1}_{\Delta_{k-1}} \left( W, \ldots, W, F \right)  
+  \sum_{\ell=1}^{\infty} 
 \left( {\scriptstyle \frac{\hbar}{i}}  \right)^{\ell} \sum_{k=0}^{\infty}\frac{1}{k!}
\Phi^{k+1}_{\Delta_{k+\ell-1}} \left( W, \ldots, W, F \right)  ~.
\eeq
The classical transformation 
\beq
 \sigma^{\rm cl}_W (F)  =  \Phi_{\Delta_0}^2(W_0,F)
\label{oribrstsym}
\eeq   
will therefore in general be replaced by
\beq
  \sum_{k=1}^{\infty}\frac{1}{k!}
\Phi^{k+1}_{\Delta_{k-1}} \left( W_0, \ldots, W_0, F \right) 
\label{defbrstsym}
\eeq

Conversely, if one assumes that quantum deformations of $\Delta$ must
be of such a form as to preserve the original classical Master Equation 
(\ref{defclasmast}), then this
places restrictions on the expansion (\ref{deltaexpand}). The condition
is that $\Delta_n$ must be a differential operator of order $\leq n+1$ 
for $n \geq 1$. 
Or, equivivalently,  $\Delta_{n,n+2}=0$ for $n \geq 1$.

\vspace{0.5cm}
\noindent
{\sc Acknowledgement:}~ The work of I.A.B.  has been partially
supported by a Human Capital and Mobility Program of the European
Community under projects INTAS 93-0633, 93-2058, and by grant No.
96-01-00482 from the Russian Foundation for Basic Researchers.
The work of K.B. and P.H.D.
has been partially supported by NorFA grants No. 95.30.182-O and 96.15.053-O, 
respectively. I.A.B. would like to thank the Niels
Bohr Institute and Prof. A. Niemi at Uppsala University for the warm
hospitality extended to him during visits there.

 \end{document}